\documentclass{article}
\usepackage{amsfonts}
\usepackage{graphicx}

\begin{document}

\date{}

\title{Vortices, Confinement\\ and Higgs fields
\footnote{\uppercase{T}alk presented by \uppercase{M}anfried \uppercase{F}aber at the 5th International Conference ``Quark confinement and the hadron spectrum'', Gargnano, Garda Lake, 10-14 September 2002. \uppercase{T}his work is supported in part by \uppercase{F}onds zur \uppercase{F}\"orderung der
\uppercase{W}issenschaftlichen \uppercase{F}orschung \uppercase{P}11387-\uppercase{PHY}.}}

\author{ROMAN BERTLE, MANFRIED FABER\\[6mm]
Atominstitut, Technische Universit\"at Wien,\\
A--1040 Vienna, Austria
\\E-mail: {bertle@kph.tuwien.ac.at, faber@kph.tuwien.ac.at}}


\maketitle

\begin{abstract}
{We review lattice evidence for the vortex mechanism of quark confinement and study the influence of charged matter fields on the vortex distribution.}
\end{abstract}

\section{Review of the vortex mechanism of quark confinement}

In the early days of QCD some people argued that quarks are confined due to the non-abelian nature of gauge-fields leading to attraction between flux-lines. But the abelian U(1) lattice gauge theory in four dimensions has also a confining phase. Therefore, it seems more reasonable to argue that the QCD vacuum is non-trivial, a condensation of topological objects which expel the color electric flux of test charges $Q$.

We have therefore to answer the question which type of large scale topological fluctuations determines the infrared behavior of the $Q\bar{Q}$-potential. To find an appropriate reply we have to specify more exactly which charges are confined in an SU(N) gauge theory. A fast first answer, SU(N) color charges, is not correct since adjoint charges are not confined, they are screened by the confining medium. Dual superconductor models derive from the confinement mechanism of U(1) that all abelian charges are confined. But we know that charges with N-ality zero are screened. Finally, we can conclude that charges with non-vanishing N-ality are confined and N-ality zero loops are not disordered. In the infrared the asymptotic string tension depends only on the N-ality of the charges.

The only known large-scale fluctuation with the required disordering of loops are vortices, magnetic flux lines which carry a flux corresponding to the center of the gauge group.

The idea that the relevant degrees of freedom are thick center vortices appeared in the late 1970's \cite{tHo78}. These spread out vortices have the topology of closed tubes in three-dimensional space and closed two-dimensional world-surfaces in four-dimensional space-time. They cost very little action and disorder the Wilson loop at large scales. By  a very simple argument the area law for N-ality $\ne 0$ Wilson loops follows from fluctuations in the number of vortices piercing the loop.

Due to the lack of an identification method for vortices almost
nothing has been done, with one notable exception \cite{Tom94}, with
the vortex idea since the early 1980's, the dawn of Monte Carlo
lattice gauge simulations. Within the dual superconductor model of
confinement appropriate methods of Abelian projection were invented
\cite{KSW87}. This led to the idea to identify vortices by similar
methods \cite{DFGO97,DFGGO98,KT98} and to develop the vortex picture into a
quantitative tool\footnote{In the following we discuss only the case of the SU(2) gauge group, although the methods were also applied to SU(3).}. It was suggested to identify vortices by maximal center
gauge in a two step process, first fixing with an over-relaxation
procedure to adjoint Landau gauge, maximizing
\begin{eqnarray}
\sum_{x,\mu}\;\Bigl| {\mathrm Tr}[U_\mu(x)] \Bigr|^2\;
\qquad \mathrm{or} \qquad
       \sum_{x,\mu}\;{\mathrm Tr}[U^A_\mu(x)]
\label{gfunct}
\end{eqnarray}
and then to project the links $U_\mu(x)$ to the nearest center element
(center projection). The physical interpretation of this procedure is
finding the best fit of the link configuration by an ensemble of thin
center vortices (P-vortices).

This method to identify vortices in Monte-Carlo configurations led to some very interesting findings concerning the confining properties:
\begin{itemize}
\item P-vortices locate thick center vortices: Vortex limited Wilson loops $W_n(C)$, the subensemble of full (unprojected) Wilson loops which are pierced by $n$ P-vortices, approach asymptotically for large loops $W_n(C)/W_0(C)=(-1)^n$. \cite{DFGGO98}

\item Center dominance: The projected string tension is close to the asymptotic string tension $\sigma$ of full lattice configurations $\chi_{cp}(R,R) \approx \sigma \quad (R \ge 2)$. \cite{DFGO97,DFGGO98,FGO00}

\item Precocious linearity: Projected Creutz ratios are approximately constant starting from small distances. \cite{DFGO97}

\item The vortex density shows asymptotic scaling. \cite{LRT98}

\item The removal of center vortices destroys confinement. \cite{DFGGO98,DD99}

\item Thick vortices lead to approximate Casimir scaling of higher-representation potentials at intermediate distances. \cite{FGO98,De00}

\item Center vortices explain screening of adjoint charges at large distances. \cite{FGO98,De00}
  
\item The deconfinement transition can be understood as percolation/depercolation transition. \cite{ELRT00,BFGO99}

\item Confinement disorder is center disorder. \cite{AG98}

\end{itemize}
\begin{figure}
\centering
\includegraphics[width=0.3\textwidth]{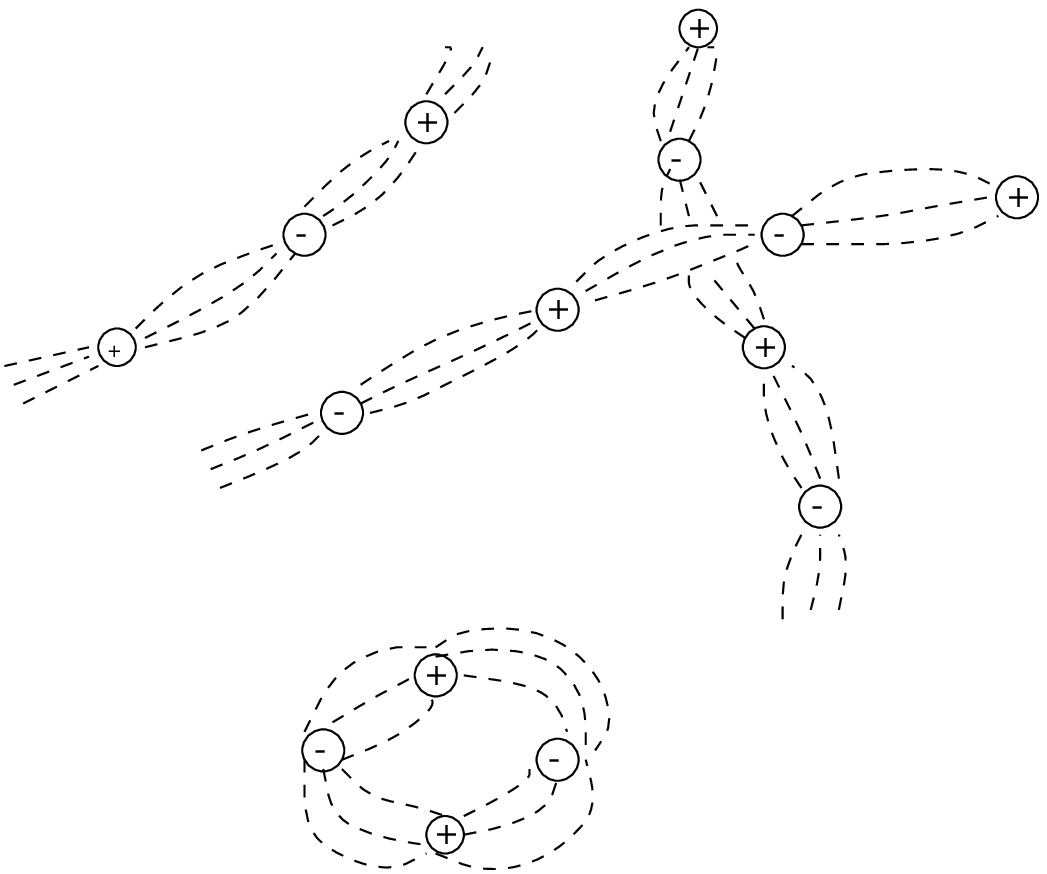}\hspace{15mm}
\includegraphics[width=0.3\textwidth]{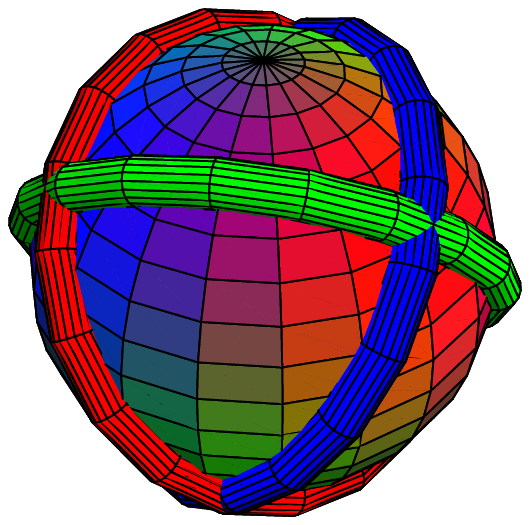}
\caption{Left: Schematic picture of a vortex as a chain of monopoles alternating with antimonopoles. Right: A thick symmetric spherical vortex after center projection (sphere), and three types of abelian monopoles (circles) after abelian projection in $\sigma_1$, $\sigma_2$ and $\sigma_3$ directions, resp.}
\label{vortmonop}
\end{figure}

By the idea that monopoles are located on vortices \cite{DFGO98} the vortex picture has some relation to the dual superconductor picture of confinement. In general vortex surfaces are composed of patches with different orientation. The lines where the orientation switches can be associated with monopole trajectories. In abelian projection vortices are chains of monopoles alternating with antimonopoles. For thick vortices covering the full gauge group SU(2), not only an abelian subgroup, it turns out that the position of monopole lines depends on the choice of the U(1) subgroup. A symmetric version of such a thick symmetric spherical vortex is depicted in Fig.~\ref{vortmonop} after center projection (sphere). The position of monopole world-lines (circles) depends on the type of abelian projection.

Vortices were originally intended to explain the confining properties
of QCD, but their presence is also correlated with topological
properties of the QCD vacuum. Removing center vortices from lattice
configurations brings them into the zero topological-charge
sector \cite{DD99}. Engelhardt and Reinhardt \cite{ER00} suggested to
determine the topological charge of a field configuration from
P-vortices by counting the contributions from singular points. This
includes self-intersections of the vortex surface and other points
with pairs of plaquettes spanning all four space-time directions, the
so called twisting points. The orientation of the vortex surface,
encoded in the sign of the field strength, enters into the definition
of the topological charge. In realistic field configurations twisting
points turn out to be far more important than self-intersection points
\cite{BEF02}. Before the topological charge can be extracted from
Monte-Carlo configurations ambiguities must be resolved which would
not appear in continuous space-time, e.g.\ vortices intersect on
lattices in general along lines. Further, UV-artifacts must be removed
by a smoothing or blocking algorithm.

Maximal center gauge has been the subject of a debate concerning large
Gribov copy effects. It turned out that the over-relaxation procedure,
used for maximizing the gauge fixing functional (\ref{gfunct}), was
more essential than originally intended \cite{FGO01} since closer
numerical scrutiny has indicated that the highest maxima of the gauge
fixing functional may not be the best \cite{KT99}. This
actually is not entirely surprising in view of the continuum limit of
the maximal center gauge \cite{ER00}, which always leads to a trivial
field configuration with no vortex content. This problem can be solved
using Laplacian center gauges \cite{AEF00}. They improve the criteria
how to get the best fit of the unprojected configuration by thin
center vortices.

A further interesting question concerns the reaction of vortices to
charged matter fields. This issue will be discussed in the next
section.


\section{P-vortices in the SU(2)-Higgs system}
In the presence of charges in the fundamental representation, i.e.\ 
with color spin $1/2$, one expects the breakdown of the confinement
potential at large distances. This screening effect and its relation
to the vortex picture can be studied for dynamical fermion fields or
for bosonic fields, as in our case. The lattice action for an SU(2)
gauge system interacting with a fundamental Higgs field is given by
\begin{eqnarray}
  S &=& S_W +
    \sum_x \left(\Phi^\dagger(x)\Phi(x) +
      \lambda\left(\Phi^\dagger(x)\Phi(x)-1\right)^2\right) -\nonumber \\[4pt]
    &&\quad\kappa \sum_{\mu,x} \left(\Phi^\dagger(x)U_\mu(x)\Phi(x+\hat{\mu})
      + \mathrm{cc}\right) \nonumber \\[4pt]
S_W &=& \beta \sum_{\mu<\nu,x} \left(1-\frac{1}{3}\mathrm{Re}\,\mathrm{Tr}\,U_{\mu\nu}(x)\right)\;,
\label{eq:action}
\end{eqnarray}
where $S_W$ is the usual Wilson plaquette action.  $\Phi$ is a complex
two-component field representing the massive scalar field in the
fundamental representation.

Screening effects by fermions were first studied by the Vienna lattice group, in the beginning of nineties \cite{FM92}. It was possible to show fermionic screening on rather small lattices at finite temperature even on a small workstation. On the other hand for many years large scale computations were not able to detect screening in zero-temperature calculations \cite{HEMCGC}. This was explained by the difficulty to detect the asymptotic time behavior of Wilson loops. G\"usken \cite{Gu98} suggested to do coupled channel calculations including the time evolution of single dynamical charges. This program was realized in Higgs models by \cite{KS98} and for fermions by \cite{Ses00}.

String breaking in the 3D Abelian Higgs model was discussed within the monopole mechanism for confinement by \cite{CIS02} and in 3D Z$_2$ gauge theory in \cite{GPP02}. In this report we study the influence of fundamental Higgs field on vortices and relate their behavior to measurements of Wilson loops and Polyakov loops. Related calculations were recently reported by Langfeld \cite{La02}.

\begin{figure}
  \centering
  \includegraphics[width=0.7\linewidth]{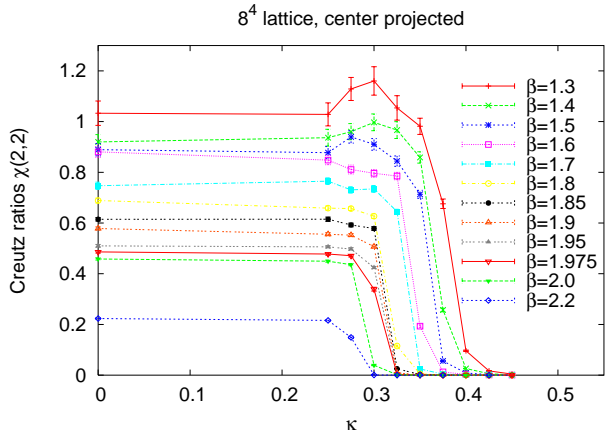}\\
  \caption{Creutz ratios $\chi(2,2)$ at zero temperature.}
  \label{creutz2L8S-1}
\end{figure}

One could expect screening of test charges in the gauge-Higgs system
if vortices split in separated pieces and vortices pierced the Wilson
loop in pairs. To get an impression of the effect of Higgs fields on
vortices we measure Wilson loops and vortex densities after center
projection in MCG fixed via over-relaxation. This method is known to
reproduce the asymptotic string tension for the pure gauge system up
to a few percent\cite{DFGO97}. In Fig. \ref{creutz2L8S-1} we show
scans of the center projected Creutz ratios $\chi(2,2)$ varying the
gauge-Higgs coupling $\kappa$ for various values of $\beta$ at $\lambda=0.5$
for an $8^4$-lattice.  The result is in agreement with measurements of
unprojected Creutz ratios: For low values of $\kappa$ the Higgs field
has little influence.  Further, projection does not affect the
critical values of $\kappa$ for the transition to the Higgs-``phase''\footnote{Since there is an analytic connection between ``confined'' and Higgs-``phase'' we put phase in quotation marks.}.
Around a $\beta$-dependent critical $\kappa$ the unprojected and
projected Creutz ratios drop rapidly to zero.

\begin{figure}
  \centering
  \includegraphics[width=0.7\linewidth]{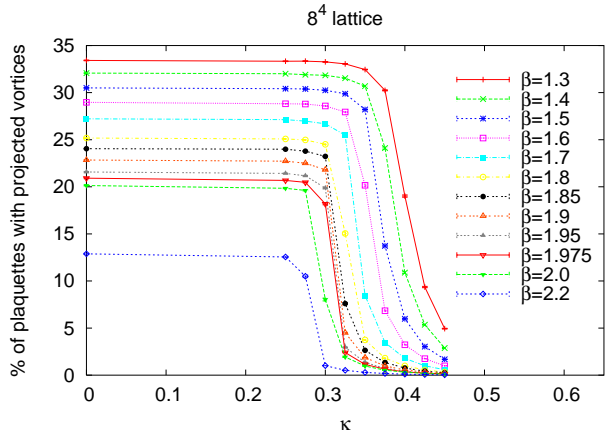}
  \caption{Percentage of plaquettes pierced by a P-vortex at zero temperature.}
  \label{FignplnegL8}
\end{figure}

Fig. \ref{FignplnegL8} shows the density of vortex plaquettes for
the same region of couplings.  Obviously, this number behaves as
expected from the measurement of Creutz ratios.  Below the transition
to the Higgs-``phase'' the density of vortex plaquettes is practically
independent of $\kappa$.  Furthermore, in this region we find only one
large vortex which extends over the whole lattice.  We see no
indication for the ``expected'' splitting of vortices.  This is in
good agreement with the results on Creutz ratios, but to explain
screening at large distances within the vortex picture another mechanism seems
required.  Such a mechanism should be compatible with the difficulties
to detect screening with Creutz ratios.  At the transition to the
Higgs-``phase'', the vortex density vanishes as do the Creutz ratios.
The vortex picture of quark confinement turns again out to be
compatible with Wilson loop measurements.

\begin{figure}
  \centering
  \includegraphics[width=0.7\linewidth]{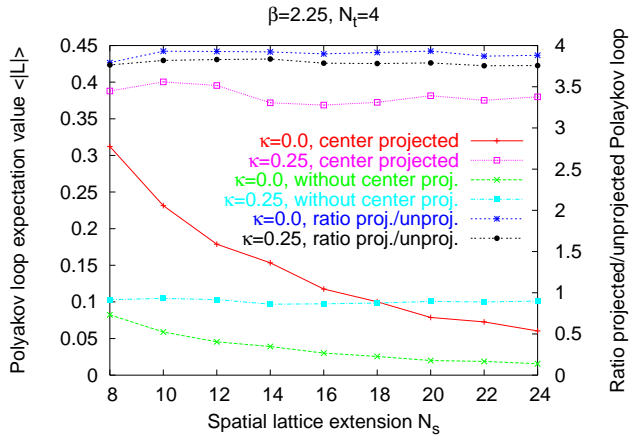}
  \caption{Expectation values (\ref{poly}) of projected and unprojected Polyakov loops. The scale for their ratio is given on the right axis.}
  \label{polyakovratioB225}
\end{figure}

As mentioned above there is a clear signal for string breaking between
static charges at finite temperature. It is interesting to see whether
string breaking can also be detected with vortices. In Fig.
\ref{polyakovratioB225} we compare the expectation values of
unprojected Polyakov loops
\begin{eqnarray}
\langle \mid L \mid \rangle 
:= \lim_{V\rightarrow \infty}\langle \mid \frac{1}{V} \sum_{\vec{r}}^V \prod_{t=1}^{N_t} U_4(\vec{r},t) \mid \rangle
= e^{-F(Q)/T}
\label{poly}
\end{eqnarray}

at $\beta=2.25$ in the pure gluonic system $\kappa=0$ and in the gauge-Higgs system at $\kappa=0.25$ below the transition to the Higgs-``phase''. In the pure gluonic system $\langle \mid L \mid \rangle$ approaches $0$ for large volumina $V$. A single test charge violates Gau{\ss}'s law and cannot exist in a confined medium. For $\kappa>0$ the test charge can be screened by fundamental Higgs charges and gets a finite free energy $F$. Lattice sizes in Fig.~\ref{polyakovratioB225} are varied between $4\times 8^3$ to $4\times 24^3$. In the same figure we show the values of the Polyakov loops which we get after MCG and center projection, see also \cite{DFGO97} for the pure gauge case. They behave in the same way as in the unprojected configuration. The absolute values of projected Polyakov loops differ from the full loops since the projection procedure removes the short range fluctuations of the gluon field. Long range physics is unchanged as one can see from the ratio of projected to unprojected Polyakov loops which is almost independent of the investigated volumina, as can also be seen in Fig.~\ref{polyakovratioB225}.

\begin{figure}
  \centering
  \includegraphics[width=0.6\linewidth]{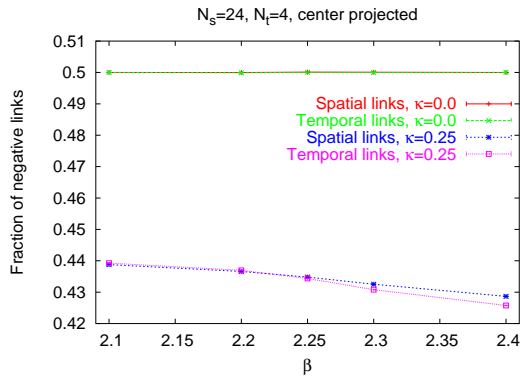}
  \caption{Fraction of negative spatial and temporal links.}
  \label{neglinks_beta}
\end{figure}

The influence of the Higgs-field on the gauge field can be discussed in unitary gauge, where the color-spin up components of the Higgs field are rotated to real positive values. The gauge-Higgs coupling leads in this gauge to a polarization of links. With increasing $\kappa$ links with positive traces dominate over those with negative traces which may lead to the finite values of $\langle \mid L \mid \rangle$. After center projection there is still a Z(2)-symmetry which is left unfixed. This gives the possibility to fix to a unitary-Z(2) gauge, where the real parts of the color-spin up components of the Higgs field is chosen positive. In this gauge the fraction of links with non-trivial center elements (negative links) is equilibrated for $\kappa=0$ and is suppressed for finite $\kappa$ as can be seen in Fig.~\ref{neglinks_beta}. This indicates that the influence of the Higgs field is not destroyed by our gauge fixing procedure.


\section{Conclusion}
One of the main aims of present day lattice investigations is to obtain a consistent picture of the QCD vacuum. During the last years, the vortex picture turned out to be a good candidate for that. In this picture the QCD vacuum is non-trivial, it is a condensate of thick random vortices with color structure. These vortices can be identified in lattice Monte-Carlo configurations in an appropriate gauge, like maximal center gauge. Center projection of the field configurations corresponds to a compression of the flux into quantised tubes. Thick vortices explain the confining properties of the QCD vacuum and there is also evidence that vortices encode the topological properties of the gauge field.

As shown in this talk the behavior of vortices is also in good agreement with the behavior of gluonic observables in the gauge-Higgs system. At zero temperature and below the transition to the Higgs-``phase'' an influence of the structure of P-vortices from the gauge-Higgs coupling could not be detected. This explains within the vortex picture the problems to prove fermionic screening with Wilson loops. At finite temperature vortices reproduce the gluonic screening of static color charges very well.


\section*{Acknowledgments}
We are grateful to Nora Brambilla and Giovanni Prosperi for the invitation to present our results at this stimulating meeting. We thank to Jeff Greensite and \v{S}tefan~Olejn{\'\i}k for the cooperation concerning the SU(2)-Higgs system and the critical reading of the manuscript, and to Michael Engelhardt for the collaboration concerning the topological susceptibility from vortices.


\begin{thebibliography}{99}

\bibitem{tHo78}G. 't~Hooft, {\it Nucl. Phys.} {\bf B138}, 1 (1978);
P.~Vinciarelli, {\it Phys. Lett.} {\bf 78B}, 485 (1978);
T.~Yoneya, {\it Nucl. Phys.} {\bf B144}, 195 (1978);
M.~Cornwall, {\it Nucl. Phys.} {\bf B157}, 392 (1979);
G.~Mack, V.~B.~Petkova, {\it Ann. Phys.} (NY) {\bf 123}, 442 (1979);
B.~Nielsen, P.~Olesen, {\it Nucl. Phys.} {\bf B160}, 380 (1979).

\bibitem{Tom94} E.~Tomboulis, {\it Nucl. Phys. Proc. Suppl.} {\bf 34}, 192 (1994); {\it Nucl. Phys. Proc. Suppl.} {\bf 30}, 549 (1993); {\it Phys. Lett}. {\bf B303}, 103 (1993).

\bibitem{KSW87} A.~Kronfeld, G.~Schierholz, U.-J.~Wiese,  {\it Nucl. Phys.} {\bf B293}, 461 (1987).

\bibitem{DFGO97} L.~Del Debbio, M.~Faber, J.~Greensite, {\v S}.~Olejn\'{\i}k, {\it Phys. Rev.}  {\bf D55}, 2298 (1997) [arXiv:hep-lat/9610005].

\bibitem{DFGGO98} L.~Del Debbio, M.~Faber, J.~Giedt, J.~Greensite, {\v S}.~Olejn\'{\i}k, {\it Phys. Rev.}  {\bf D58}, 094501 (1998) [arXiv:hep-lat/9801027].

\bibitem{KT98} T.~G.~Kov\'{a}cs, E.~T.~Tomboulis, {\it Phys. Rev.}  {\bf D57}, 4054 (1998) [arXiv:hep-lat/9711009].

\bibitem{FGO00}
M.~Faber, J.~Greensite, {\v S}.~Olejn\'{\i}k, {\it Phys. Lett.}  {\bf B474}, 177 (2000) [arXiv:hep-lat/9911006].

\bibitem{LRT98}
K.~Langfeld, H.~Reinhardt, O.~Tennert, {\it Phys. Lett.}  {\bf B419}, 317 (1998) [arXiv:hep-lat/9710068].

\bibitem{DD99} 
Ph.~de~Forcrand, M.~D'Elia, {\it Phys. Rev. Lett.}  {\bf 82}, 4582 (1999) [arXiv:hep-lat/9901020].

\bibitem{FGO98}
M.~Faber, J.~Greensite, {\v S}.~Olejn\'{\i}k, {\it Phys. Rev.}  {\bf D57}, 2603 (1998) [arXiv:hep-lat/9710039].

\bibitem{De00}
S.~Deldar,  {\it JHEP}  {\bf 0101}, 013 (2001) [arXiv:hep-ph/9912428].

\bibitem{ELRT00} M.~Engelhardt, K.~Langfeld, H.~Reinhardt, O.~Tennert, {\it Phys. Rev.} {\bf D61}, 054504 (2000) [arXiv:hep-lat/9904004].

\bibitem{BFGO99} R.~Bertle, M.~Faber, J.~Greensite, {\v S}.~Olejn\'{\i}k, {\it JHEP} {\bf 03}, 019 (1999) [arXiv:hep-lat/9903023].

\bibitem{AG98} J.~Ambj{\o}rn, J.~Greensite, {\it JHEP} {\bf05}, 004 (1998) [arXiv:hep-lat/9804022].

\bibitem{DFGO98} L.~Del Debbio, M.~Faber, J.~Greensite, {\v
  S}.~Olejn\'{\i}k, in {\em New Developments in Quantum Field
  Theory}, ed.  {Poul~Henrik Damgaard and Jerzy Jurkiewicz} (Plenum
  Press, New York--London, 1998) 47 [arXiv:hep-lat/9708023];
J.~Ambj{\o}rn, J.~Giedt, J.~Greensite, {\it JHEP}  {\bf0002}, 033 (2000) [arXiv:hep-lat/9907021].

\bibitem{ER00} M.~Engelhardt, H.~Reinhardt, {\it Nucl. Phys.} {\bf B567}, 249 (2000) [arXiv:hep-th/9907139].

\bibitem{BEF02} R.~Bertle, M.~Engelhardt, M.~Faber, {\it Phys. Rev.}  {\bf D64}, 074504-1-10 (2001) [arXiv:hep-lat/0104004].

\bibitem{FGO01} M.~Faber, J.~Greensite, \v{S}.~Olejn{\'\i}k, {\it Phys. Rev.}  {\bf D64}, 034511 (2001) [arXiv:hep-lat/0103030].

\bibitem{KT99} T.~G.~Kov\'{a}cs, E.~T.~Tomboulis, {\it Phys. Lett.} {\bf B463}, 104 (1999) [arXiv:hep-lat/9905029];
V.~G.~Bornyakov, D.~A.~Komarov, M.~I.~Polikarpov,
A.~I.~Veselov, {\it JETP Lett.} {\bf 71}, 231 (2000) [arXiv:hep-lat/0002017];
R.~Bertle, M.~Faber, J.~Greensite, \v{S}.~Olejn{\'\i}k,
 {\it JHEP} {\bf 0010}, 007 (2000) [arXiv:hep-lat/0007043];
V.~G.~Bornyakov, D.~A.~Komarov, M.~I.~Polikarpov, {\it Phys. Lett.} {\bf B497}, 151 (2001) [arXiv:hep-lat/0009035].

\bibitem{AEF00} C.~Alexandrou, M.~D'Elia, P.~de~Forcrand,
{\it Nucl. Phys. Proc. Suppl.} {\bf 83}, (2000) 437 [arXiv:hep-lat/9907028];
C.~Alexandrou, P.~de~Forcrand, M.~D'Elia,
{\it Nucl. Phys.} {\bf A663}, 1031 (2000) [arXiv:hep-lat/9909005];
P.~de~Forcrand, M.~Pepe,
{\it Nucl. Phys.} {\bf B598}, 557 (2001) [arXiv:hep-lat/0008016];
M.~Faber, J.~Greensite, \v{S}.~Olejn{\'\i}k, {\it JHEP} {\bf 0011}, 012 (2001) [arXiv:hep-lat/0106017];
K.~Langfeld, H.~Reinhardt, A.~Sch\"afke, {\it Phys. Lett.} {\bf B504}, 338 (2001) [arXiv:hep-lat/0101010].

\bibitem{FM92} W.~Feilmair, H.~Markum, {\it Nucl. Phys.} {\bf B370}, 299 (1992).

\bibitem{HEMCGC} U.~M.~Heller, K.~M.~Bitar, R.~G.~Edwards, A.~D.~Kennedy [HEMCGC Collaboration],
{\it Phys. Lett.} {\bf B335}, 71 (1994) [arXiv:hep-lat/9401025].

\bibitem{Gu98} S.~Güsken, {\it Nucl. Phys. Proc. Suppl.} {\bf 63}, 16 (1998).

\bibitem{KS98} F.~Knechtli, R.~Sommer [ALPHA Collaboration],
 {\it Phys. Lett.}  {\bf B440}, 345 (1998) [arXiv:hep-lat/9807022];
O.~Philipsen, H.~Wittig, {\it Phys. Rev. Lett.}  {\bf 81}, 4056 (1999) [Erratum-ibid.  {\bf 83}, 2684 (1999)] [arXiv:hep-lat/9807020];
P.~W.~Stephenson, {\it Nucl. Phys.}  {\bf B550}, 427 (1999) [arXiv:hep-lat/9902002].

\bibitem{Ses00} B.~Bolder {\it et al.} [SESAM Collaboration], {\it Phys. Rev.} {\bf D63}, 074504 (2001) [arXiv:hep-lat/0005018];
C.~W.~Bernard {\it et al.}, {\it Phys. Rev.} {\bf D64}, 074509 (2001) [arXiv:hep-lat/0103012].

\bibitem{CIS02}M.~N.~Chernodub, E.~M.~Ilgenfritz, A.~Schiller, {\it Phys. Lett.} {\bf B547}, 269 (2002) [arXiv:hep-lat/0207020].

\bibitem{GPP02} F.~Gliozzi, M.~Panero, P.~Provero, {\it Phys. Rev.} {\bf D66}, 017501 (2002) [arXiv:hep-lat/0204030].

\bibitem{La02} K.~Langfeld, Talk at the International conference on Strong and electro-weak matter; Heidelberg, 2002 [arXiv:hep-lat/0212032].

\end{thebibliography}
\end{document}